\documentclass{aa} % ajouter referee dans les crochets pour passer à  1 colonne. Enlever pour 2 colonnes
\usepackage{graphicx}
\usepackage{subfigure}
\usepackage{color}
\usepackage{comment}
\usepackage{amsmath}
\usepackage{breqn}
\usepackage{ulem}
\usepackage{amsfonts}
\usepackage{amssymb}
\usepackage{booktabs}
\usepackage{inputenc}
\usepackage[T1]{fontenc}
\usepackage{mathrsfs}
\usepackage{enumerate}
\usepackage{xcolor,url}
\usepackage{cancel,dsfont} 
\usepackage{multirow}
\usepackage{soul}
\usepackage{braket}
\usepackage{natbib}
\usepackage{float}
\usepackage[varg]{txfonts}

\usepackage{subcaption}
\usepackage{subfig}

\begin{document}

\title{
21cm signal from dark-age collapsing halos with a detailed molecular cooling treatment
}

\titlerunning{21cm signal from dark-age collapsing halos with detailed molecular cooling treatment}
\author{Hugo Plombat\inst{1}
           , Denis Puy\inst{1}}
\authorrunning{H. Plombat, D. Puy}          

\institute{\inst{1}$\,$Laboratoire Univers et Particules de Montpellier (LUPM), CNRS \& IN2P3 et Universit\'e de Montpellier \\ (UMR-5299),
 Place Eug\`ene Bataillon, F-34095 Montpellier Cedex 05, France \\
 \email{hugo.plombat@umontpellier.fr, denis.puy@umontpellier.fr}}

\date{Accepted April 23 2025 }

\long\def\Htwo{H$_2$}

\abstract{
\vskip2mm
\noindent
\textit{Context}. 
 To understand the formation of the first stars, a detailed description of the thermal and chemical processes in collapsing gas clouds is essential. Molecular cooling, particularly via $ \rm H_2$, plays a significant role in triggering thermal instabilities that lead to star formation. The 21cm hydrogen line serves as a potential probe of the first collapsing structures during the dark ages of the early Universe and it is affected by the gas temperature evolution.

\vskip2mm
\noindent
\textit{Aims}. We aim to investigate the molecular cooling in the gas halos prior to the formation of the first stars, with a particular focus on how the $ \rm H_2$ cooling affects the gas temperature. Additionally, we explore the sensitivity of the 21cm hydrogen line to these cooling processes during the collapse of the first overdense regions.
\vskip2mm
\noindent
\textit{Results}. 
We introduce the \texttt{CHEMFAST} code, which tracks the evolution of chemical abundances and computes the 21cm neutral hydrogen signal in collapsing halos. Our results show that molecular cooling significantly affects the gas temperature inside collapsing clouds of mass ranging from $10^6$ to $10^9$ M$_\odot$, influencing the 21cm signal. The signal exhibits an emission feature that is distinct from the one predicted in simpler expansion models.

\vskip2mm
\noindent
\textit{Conclusions}. 
 The 21cm brightness temperature inside collapsing clouds displays an emission feature driven by molecular cooling, closely mirroring the gas temperature evolution. This makes the dark-age 21cm signal a promising probe for studying the thermal processes and structure formation in the early Universe.
}

\maketitle

\section{Introduction}

The chemistry of the early Universe has been the target of large studies aimed at determining the abundances of elements and dedicated to the calculation of reaction rates that govern chemical processes, which are contingent on the prevailing environmental conditions. Such investigations explore the primordial species produced by the first nuclear reactions during standard Big Bang nucleosynthesis (SBBN), (see e.g., ~\cite{particledatagroup22}, \cite{signorepuy09}, \cite{cyburt16}, \cite{pitrou21}, \cite{schoneberg24}), up to the interstellar medium (ISM), hosting a variety of heavier elements forged within stars \citep{tielens05,herbst95}.  \\

From the SBBN, the Universe was completely ionized, as the temperature of the baryon-photon fluid, maintained by interactions via Compton scattering between free electrons and photons, was too high for any neutral atom to form. However, as the Universe expanded, the effectiveness of baryon-radiation coupling diminished, allowing ion-electron reactions to dominate. Consequently, this led to the progressive recombination of ionized elemental forms into their neutral counterparts \citep{zeldovich68,peebles68}. Simultaneously, despite their much lower reaction rates due to the necessarily high amount of energy for these processes to happen, molecules also began to form \citep{puy93}. \\ 

Species abundance estimation is a central problem of any coupled reaction network. One of the first detailed models of molecular synthesis were proposed by \cite{solomon72}, who computed molecular abundances in diffuse interstellar clouds, and by \cite{herbst73} who studied the formation of molecules in dense dark clouds. The article of \cite{prasad80}, which gave a comprehensive library consisting of over 1400 reactions for 137 species, presented a first model for gas phase chemistry in interstellar clouds. The UMIST Database~\citep{leteuff00,mcelroy13}, contains the updated rate coefficient, temperature ranges and temperature dependence of 6173 gas-phase reactions important in astrophysical environments.  However at early epochs, when a total absence of dust grains appears justified, the chemistry is different from the typical interstellar medium chemistry. From the exhaustive work of~\cite{bates51} on the radiative association rate,~\cite{takayanagi60,hirasawa69,takeda69,saslaw67} and \cite{matsuda69} pointed out the way of H$_2$ formation through H$_2^+$ and H$^-$ without grains in the early Universe. Several chemical networks studies including the primordial molecules (such as H$_2$, HD, and LiH) and ions were carried out by \cite{lepp84,latter91,puy93,Galli:1998dh,stancil98a, lepp02, Pfenniger:2002ji,hirata06,glover08,signorepuy09,coppola11,gay11,coppola12,coppola13}. \\ 

The determination of the molecular abundances is essential in the study of the formation of structures. Indeed, the excitation of the rotational and vibrational levels of the molecules inside collapsing matter overdensities gives rise to a molecular cooling mechanism, which can induce the appearance of instabilities of a thermal nature, highlighted for the first time in  \cite{field65}  and later in \cite{yoshii80}. This excitation is necessary in order to trigger the formation of the first stars~\cite{Tegmark:1996yt,puy97,Yoshida:2006bz,Trenti:2009cj,galli13,Bromm:2013iya,Glover:2012gx, ripamonti04}. Molecular cooling has been extensively studied for numerous molecular species, the most important being $ \rm H_2$, which despite the absence of a dipole moment is the stronger coolant due to its  high abundance \citep{abel97,Galli:1998dh, stancil98b, glover08}. $\rm HD$ is also expected to be an efficient coolant due to its non-zero dipole moment, but it has been shown that it only dominates the molecular cooling at very low temperatures ($\sim 150 \rm K$; \citealt{glover08}), which are typically not reached in primordial gas during the collapse of the first overdense regions \citep{bromm02}.
Other early molecules, such as $ \rm H_2^+$ \citep{yoshida07,glover09}, $ \rm H_3^+$  \citep{glover06,glover09}, and LiH \citep{stancil96, mizusawa05}, have minor cooling contributions in the context of setting up the conditions for the formation of the first stars. \\

Probing the formation of the first structures preceding star formation, during the so-called dark ages, is very challenging because almost no signals are emitted from this epoch. Baryonic matter predominantly existed in a neutral state, as the Compton coupling with radiation waned with cosmic expansion.  The 21cm hydrogen line, arising from hyperfine spin-flip transitions in neutral hydrogen atoms, offers a promising avenue for probing this epoch. \\

Many future experiments are planned in order to measure the 21cm signal during  cosmic dawn and the Epoch of Reionization, such as Experiment to Detect the Global rEionization Signature (EDGES; \citealt{bowman18}), Shaped Antenna measurement of the background RAdio Spectrum \citep[SARAS;][]{nambissan21}, Large-Aperture Experiment to Detect the Dark Ages \citep[LEDA;][]{price18}, the Radio Experiment for the Analysis of Cosmic Hydrogen \citep[REACH;][]{deleraacedo22}, and Probing Radio Intensity at high-Z from Marion \citep[PRizM;][]{philip19}. These studies are aimed at the observation of the 21cm global signal, whereas ground based interferometers such as the Hydrogen Epoch of Reionization Array \citep[HERA;][]{DeBoer:2016tnn}, Low Frequency Array \citep[LOFAR;][]{vanhaarlem13, patil16}, Murchison Widefield Array \citep[MWA;][]{bowman13,tingay13,jacobs16}, the Giant Metrewave Radio Telescope~\citep[GMRT;][]{swarup91}, the Precision Array for Probing the Epoch of Reionization \citep[PAPER;][]{parsons10}, and the future Square Kilometer Array \citep[SKA;][]{koopmans15} are measuring or planning to measure  the spatial fluctuations in the signal.  \\ 

However, the observation of the  redshifted 21cm power spectrum and global signal from the dark ages is extremely challenging and cannot be done from Earth-based experiments, because of the ionosphere which acts as a highly opaque foreground for the frequencies of interest, as well as the radio frequency interferences of human origin. 
The greatest hopes to fathom the dark ages are placed on space missions, either in orbit around the Moon, with the satellites and nanosatellites. We also refer to \cite{burns21b} and \cite{fialkov24},  DSL \citep{shi22}, NCDL \citep{bentum20}, and CosmoCube \citep{artuc24}, or directly on the Moon's surface with ROLSES\cite{burns21}, LuSEE-Night \citep{bale23}, the Lunar Crater Radio Telescope\citep{goel22}, FarView \citep{polidan24}, the Astronomical Lunar Observatory \citep[ALO;][]{kleinwolt24}, FARSIDE \citep{burns21}, and LARAF \citep{chen24}.  A mission  with similar objectives, known as Primordial Hydrogen Observation with Nanosatellites Explorer (PHONE), is also in co-development by the Laboratoire Univers et Particules de Montpellier (LUPM) and the Spatial Center, dedicated to the development of nanosatellites, at the  University of Montpellier. \\

Our work takes part in the studies of the 21cm signal from minihalos. However, unlike aforementioned works, which concentrate on estimating the signal from halos that have already been virialized, we focus on the earlier stages, from the turnaround point of an overdense region, to the end of its collapse. In particular, we aim to highlight the impact of molecular cooling on the 21cm signal arising from the halos. \\

The originality of this work lies in coupling the evolution of interactions between atomic and molecular ions, neutral atoms, and molecules with the dynamic and thermal evolution of a medium undergoing gravitational collapse. While this context has been extensively studied in the past to understand the fragmentation of a molecular cloud by the process of thermal instability, in this article, we integrated the calculation of the emissivity of the excitation process of neutral hydrogen producing a hyperfine transition and followed “step by step” this emissivity during a collapse. This coupled approach enables us to predict the emissivity of the 21 cm line at high redshifts. This work is part of a large body of research into the production of the 21 cm line during the cosmological dark ages.\\

Numerous studies have been carried out to characterize the first structures with 21cm hydrogen line, before the formation of the first stars.
\cite{iliev02}  proposed  a model of 21cm emission in minihalos from collisional excitation, before reionization.
\cite{meiksin11} followed spherical collapsing density perturbations,  estimating the 21cm signature arising from the fully collapsed minihalos accounting for the effects of molecular hydrogen formation. A statistical 21cm signal from the halo has been provided,  demonstrating how this signature could distinguish predictions of the small scales of the 21cm power spectrum. \cite{furugori20} calculated the 21cm signal emitted by ultracompact minihalos formed at high redshift. \cite{novosyadlyj20} calculated the 21cm emission in halos between $10^6$ and $10^{10} \rm M_{\odot}$, virialized between $z \sim 10 - 50$. \\  
Recently, \cite{novosyadlyj24b} estimated the  global signal in the redshifted hyperfine structure line 21 cm of hydrogen atoms formed during the dark ages and cosmic dawn. They show that the profile of this line crucially depends on the temperature and ionization of baryonic matter as well as the spectral energy distribution of radiation from the first sources. \\

In this paper, we make use of the code \texttt{CHEMFAST} initiated by \cite{puy93,Pfenniger:2002ji} and \cite{vonlanthen09}. The original purpose of this code is to compute the abundances of atomic and molecular species in the context of cosmological expansion of the universe. We incorporated the computation of the global 21cm line signal arising from collisional coupling during the dark ages. Then, by changing the equations of dynamics in the code, we followed a collapse scenario of a $10^6$-$10^9$ M$_\odot$ overdense region, modeling the region with a simple homogeneous spherical model, including pressure effects \citep{lahav86,puy96}. We emphasized the cooling effects on the gas temperature, arising from molecular excitation during the collapse. We additionally computed the 21cm signal from the forming protocloud, highlighting its distinct features compared to the expected sky-averaged 21cm global signal. \\

This paper is organized as follows. In Sect. \ref{sec:homogeneous}, we start by developing the reaction network and equations of dynamics in the context of expanding homogeneous Universe. We present the \texttt{CHEMFAST} code computation of atomic and molecular abundances. In Sect. \ref{sect:21cm}, we introduce the physics of the 21cm global signal during the dark ages.
In Sect. \ref{sect:collapse}, we describe the dynamics of the spherical collapse model and, finally, we present  our analysis of the thermal and 21cm signal evolution in collapsing halos in Sect. \ref{sect:results}. 
We present our conclusions and outline our future research directions in Sect. \ref{sect:conclusion}. Throughout this paper, we adopted the $\Lambda \rm CDM$ cosmology, using the parameters values from Planck2018 \cite{Planck:2018vyg} TT,TE,EE+lowE+lensing best-fit.

\section{\texttt{CHEMFAST} in homogeneous universe}
\label{sec:homogeneous}

In this section, we consider the well-studied case of the background evolution of chemistry during the recombination and the dark ages. Our aim is to present how the code works  using this example, and to introduce the concepts that will also be employed in this study. \\

In \texttt{CHEMFAST,} we follow a system of coupled differential equations in order to track the evolution of species abundances.  
The  SBBN model for the Universe predicts the nuclei abundances of mainly hydrogen, helium, and lithium as well as their isotopes. The chemistry of the early Universe is the chemistry of these light elements and their respective isotopic forms.
The ongoing physical reactions are numerous after the nucleus recombination. These reactions fall into three distinct categories:

\begin{itemize}
\item Collisional reactions $\xi + \xi ' \longleftrightarrow \xi_1 + \xi_2$\\
involve processes such as association, associative detachment, mutual neutralization, charge exchange, and reverse reactions; 
\item Electronic reactions $\xi + {\rm e}^- \, \longleftrightarrow \xi_1 + \xi_2$\\
encompass recombination, radiative attachment, and dissociative attachment processes;
\item Photo-processes $\xi +\gamma \, \longleftrightarrow \xi_1 + \xi_2$\\
include dissociation, detachment, ionization, and radiative association reactions, where $\gamma$ is a photon.
\end{itemize}
The chemical kinetic of a reactant, $\xi$, which leads to the products $\xi_1$ and $\xi_2$, imposes the following evolution
of the average number density $\overline{n}_{\xi}$ (number of species $\xi$ per cm$^{3}$ in the homogeneous Universe, i.e., the mean density):
\begin{equation}
\Bigr( \frac{{\rm d}\overline{n}_\xi}{{\rm d}t} \Bigl)_{\rm chem} \, = \, \sum_{\xi_1 \xi_2} k_{\xi_1 \xi_2} \, \overline{n}_{\xi_1}
\overline{n}_{\xi_2}
- \sum_{\xi '}k_{\xi \xi '} \overline{n}_\xi \overline{n}_{\xi '}.
\label{eq:kin}
\end{equation}
Here, $k_{\xi_1 \xi_2}$ represents the reaction rate of the $\xi$-formation process from $\xi_1$ and $\xi_2$, and $k_{\xi \xi '}$ denotes the reaction rate of $\xi$-destruction by collision with the reactant, $\xi '$.
The typical rate of collisional reactions, denoted as $k_{\rm coll}$ (in cm$^3$ s$^{-1}$), is calculated by averaging cross-sections $\sigma(E)$ over a Maxwellian velocity distribution at temperature, $T_{\rm k}$:
\begin{equation}
k_{\rm \xi,\xi'} = \sqrt{\frac{8}{\pi m_{\rm r}}} \, \Bigr[ k_{\rm B} T_{\rm k} \Bigl] ^{-3/2}
\int_0^\infty  \sigma(E) \, {\rm e}^{-E/k_B T_{\rm k}} \, E \, {\rm d}E
,\end{equation}
where $E$ is the collision energy, $m_{\rm r}$ the reduced mass of the collisional system ($\xi,\xi'$), $T_{\rm k}$ the kinetic baryon temperature,
and $k_{\rm B}$ the Boltzmann constant.\\
The radiative rate coefficient, $k_{\rm rad}$ (in s$^{-1}$), which depends on the radiative cross-section, $\sigma_{\rm rad}$, at the frequency, $\nu$, and
on the CMB, is defined by
\begin{equation}
k_{\rm rad} = \frac{8\pi}{c^2} \, \int_{\nu_{\rm th}} ^\infty \sigma_{\rm rad}(\nu) \,
\frac{\nu^2 \, {\rm d}\nu} {{\rm exp}(h_{\rm P}\nu / k_{\rm B} T_{\rm r}) -1 }
,\end{equation}
where $\nu_{\rm th}$ is the threshold frequency above which the radiative process is possible, $T_{\rm r}$ is the radiation
temperature, $c$ is the speed of light, and $h_{\rm P}$ is the Planck's constant.

\subsection{Reaction network}
The chemical network, namely, every reaction between atoms, ions, and molecules that we take into account in  the species abundances computation, is described in Table~\ref{tab:chemnet}, along with a summary of the reactions and their references. The primordial chemistry of heavier element is also implemented in \texttt{CHEMFAST} (Li, C, N, O, and F)~\cite{vonlanthen09}, but we ignored it as it is irrelevant for this work.
The fitting formula of the reaction rates were mainly taken from  the compilation given by \cite{schleicher08}, except for the hydrogen  and deuterium recombination, which were calculated from the rate given by \cite{novosyadlyj22}. 

\begin{table*}[ht]
\caption{\texttt{CHEMFAST} chemical network}
\centering
\scalebox{0.9}{
\begin{tabular}{| c c c |c c c|} 
 \hline
 & Reaction & Reference & & Reaction & Reference \\
 \hline
 (H1) & $\rm{H^+ + e^- \rightarrow H + \gamma} $ & NKMS22 & (H2) & $\rm{H + \gamma \rightarrow H^  + e^- }$ & AAZ97  \\
 (H3) & $\rm{H+e^- \rightarrow H^- + \gamma}$ & GP98 & (H4) & $\rm{H^- + \gamma \rightarrow H + e^-} $ & GP98 \\
 (H5) & $\rm{H^- + H \rightarrow H_2 + e^-} $ & GP98 & (H6) & $\rm{H^- + H^+ \rightarrow H_2^+ + e^-} $ & GP98 \\
 (H7) & $\rm{H^- + H^+\rightarrow H + H }$ & LSD02 & (H8) & $\rm{H + H^+ \rightarrow H_2^+ + \gamma} $ & GP98 \\
 (H9) & $\rm{H_2^+ + \gamma \rightarrow H + H^+} $ & GP98 & (H10) & $\rm{H_2^+ + H \rightarrow H_2 + H^+} $ & GP98 \\
 (H11) & $\rm{H_2^+ +  e^- \rightarrow H + H} $ & GP98 & (H12) & $\rm{H_2^+ + \gamma \rightarrow H^+ + H^+ + e^-} $ & GP98 \\
 (H13) & $\rm{H_2 + H^+ \rightarrow H_2^+ + H} $ & SKHS04 & (H14) & $\rm{H_2 + e^-\rightarrow H + H^-} $ & CCDL07 \\
 (H15) & $\rm{H_2 + e^- \rightarrow H + H + e^-} $ & TT02 & (H16) & $\rm{H_2 + \gamma \rightarrow H_2^+ + e^-} $ & GP98 \\
 (H17) &$\rm{H + H + H \rightarrow H_2 + H}$      & GL08  & (H18) & $\rm{H^- + H_2^+\rightarrow H_2 + H} $ & AAZN97   \\
 \hline    
 (D1) & $\rm{D^+ + e^- \rightarrow D + \gamma} $ & NKMS22 & (D2) & $\rm{D + \gamma \rightarrow D^+ + e^-} $ & AAZ97 \\
 (D3) & $\rm{D+ H^+ \rightarrow D^+ + H} $ & Savin (2002) & (D4) & $\rm{D^+ + H\rightarrow D + H^+} $ & Savin (2002)\\   
 (D5) & $\rm{H + D \rightarrow HD + \gamma} $ & GP98 & (D6) & $\rm{D + H_2 \rightarrow H + HD} $ & GP02 \\
 (D7) & $\rm{HD^+ + H \rightarrow H^+ + HD} $ & SLD98 & (D8) & $\rm{D^+ + H_2 \rightarrow H^+ + HD} $ & GP02 \\
 (D9) & $\rm{HD + H\rightarrow H_2 + D} $ & GP02 & (D10) & $\rm{HD + H^+ \rightarrow H_2 + D^+} $ & GP02 \\
 (D11) & $\rm{D  + H^+ \rightarrow HD^+ + \gamma} $ & GP98 & (D12) & $\rm{H + D^+ \rightarrow HD^+ + \gamma } $ & GP98 \\
 (D13) & $\rm{HD^+  + \gamma \rightarrow H  + D^+} $ & SGP+08 & (D14) & $\rm{HD^+ + \gamma \rightarrow H^+ + D} $ & SGP+08 \\
 (D15) & $\rm{HD^+ + e^- \rightarrow H + D} $ & SLD98 & &&\\
 \hline
 (He1) & $\rm{He^{2+} + e^- \rightarrow He^+ + \gamma} $ & GP98 & (He2) & $\rm{He^+ + \gamma \rightarrow He^{2+} + e^-} $ & GP98 \\
 (He3) & $\rm{He^{+} + e^- \rightarrow He + \gamma} $ & GP98 & (He4) & $\rm{He + \gamma \rightarrow He^{+} + e^-} $ & GP98\\
 (He5) & $\rm{He + H^+ \rightarrow He^+ + H} $ & GP98 \& GJ07 & (He6) & $\rm{He^+ + H \rightarrow He + H^+} $ & ZDKL89 \\
 (He7) & $\rm{He^+ + H \rightarrow HeH^+ + \gamma}$\footnote{radiative association} & SLD98 &  (He8) & $\rm{He^+ + H  + \gamma \rightarrow HeH^+ + \gamma}$\footnote{stimulated radiative association}   & ZSD98 \\
 (He9)& $\rm{He + H_2^+ \rightarrow HeH^+ +H}$ & GP98  & (He10) & $\rm{He + H^+ \rightarrow HeH^+ + \gamma}$ & SLD98 \\
 (He11) & $\rm{HeH^+ + e^- \rightarrow He + H_2^+}$ & LJB95 & (He12) & $\rm{HeH^+ + H_2 \rightarrow He +H}$ & SLD98 \\
 (He13) & $\rm{HeH^+ + \gamma \rightarrow He + H^+}$ & JSK95  & (He14) & $\rm{HeH^+ + \gamma \rightarrow He^+ + H}$ & GP98 \\
 \hline
\end{tabular}
}\\
\vspace{0.5em}
\noindent\textbf{References: } AAZ97: \cite{abel97}, CCDL07: \cite{capitelli07}, GLO08: \cite{glover08}, GP98:  \cite{Galli:1998dh}, GP02: \cite{Galli:2002du} , JSK95: \cite{jurek95}, LJB95: \cite{linder95} LSD02 :\cite{lepp02} , NKMS22: \cite{novosyadlyj22},  SGP+08: \cite{schleicher08}, SKHS04: \cite{savin04},  SLD98: \citeyear{stancil98b}, ZDKL89: \cite{zygelman98}, ZSD98: \cite{zygelman98}.
\label{tab:chemnet}

\end{table*}

\subsection{Density and temperature evolution}  

The complete system of differential equations allows us to calculate the abundances of species. We also followed the thermal evolution of the Universe (gas and radiation temperature), which plays a role in the reaction rates.

\paragraph{\textbf{Baryon number density}}

The mean baryon number density, $\overline{n}_{\rm b}$, is expressed by
\begin{equation}
\overline{n}_{\rm b} \, = \, \sum_{\xi} \, \overline{n}_\xi \, 
,\end{equation}

summed over  $\xi$, the species involved in the network. Thus, the evolution of the mean number density is given by
\begin{equation}
\frac{{\rm d}\overline{n}_{\rm b}}{{\rm d}t} \, = \, -3 H(t) \overline{n}_{\rm b} + \sum_\xi \Bigr( \frac{{\rm d}\overline{n}_\xi}{{\rm d}t} \Bigl)_{\rm chem}
\label{eq:dens_exp}
,\end{equation}
where $H(t)=\dot{a}/{a}$ is the Hubble parameter. The second term on the right-hand side of this equation accounts for the variation of the number of particles due to the chemical reactions, as detailed in  Eq.~\ref{eq:kin}.

\paragraph{\textbf{Species densities}}

For each species, $\xi$, we have
\begin{equation}
\frac{{\rm d}\overline{n}_\xi}{{\rm d}t} = - 3 H(t) \, \overline{n}_\xi
+ \Bigr( \frac{{\rm d}\overline{n}_\xi}{{\rm d}t} \Bigl)_{\rm chem}
\label{eq:denschem_exp}
,\end{equation}
where $\overline{n}_\xi$ is the number density of the species, $\xi$. The last term of the second member of this equation expresses the contribution of the chemical kinetics (see Eq.~\ref{eq:kin}).

\paragraph{\textbf{Radiation temperature}}

In an expanding universe, the energy density of radiation evolves as $\rho_r \propto (1+z)^4$. The CMB can be likened to a blackbody, and Stefan-Boltzmann's law can be used to relate its temperature to its energy density: $\rho_r \propto T_\gamma^4$.  This leads to the following equation $T_\gamma(z)= T_{\gamma,0}(1+z)$, where $T_{\gamma,0} = 2.726 \rm K$ \cite{cobe94} is the blackbody temperature of the CMB.
We can then deduce the evolution law: 
\begin{equation}
    \frac{dT_\gamma}{dt} = -H(t)T_\gamma
    \label{eq:trad_exp}
\end{equation}

\paragraph{\textbf{Gas temperature}}

The evolution of gas kinetic temperature $T_k$ is driven by three thermal channels:
\begin{equation}
\frac{{\rm d}T_{\rm k}}{{\rm d}t} \, = \, \frac{2}{3\overline{n}_{\rm b} k_{\rm B}}\Bigr(\Psi_{\rm adia} + \Psi_{\rm com} + \Psi_{\rm mol} 
\Bigl);
\label{eq:tkin_exp}
\end{equation}
 $\Psi_{\rm adia}$ is the adiabatic cooling due to the expansion is given by 
\begin{equation}
\Psi_{\rm adia} \, = \, -3H(t) n_{\rm b}k_{\rm B} T_{\rm k};
\end{equation}
$\Psi_{\rm com}$ is the energy transfer between matter and radiation via Compton
scattering of CMB photons on free electrons \citep{weymann65,peebles68};
\begin{equation}
\Psi_{\rm com} \, = \, 4 \overline{n}_{\rm e} \sigma_{\rm T} a_{\rm r} k_{\rm B} T_{\rm r}^4\, \frac{T_{\rm r} - T_{\rm k}}{m_{\rm e} c} \, .
\label{eq:compton}
\end{equation}
Here, $\sigma_T$ is the Thomson cross section, $a_r$
is the radiation constant,  $m_{\rm e}$ the electron mass, and $\overline{n}_{\rm e}$ the mean electron number density. $\Psi_{\rm mol}$ defines the energy transfer via excitation and de-excitation of molecular rotational transition. This process is discussed in Sect. \ref{sec:thermal}.\\

Finally,  the last equation  which translates the conversion between time and redshift dependence completes the system of equations: 
\begin{equation}
    \frac{dz}{dt} = -H(z) (1+z)
    \label{eq:redshift}
.\end{equation}
 
\noindent Equations \ref{eq:dens_exp}, \ref{eq:denschem_exp}, \ref{eq:trad_exp}, \ref{eq:tkin_exp}, and \ref{eq:redshift} are coupled and must be solved simultaneously.

\subsection{Numerical approach and initial conditions}
The search for solutions for systems of ordinary differential equations is a typical stiff problem. Gear methods have excellent stability properties and are widely used for solving chemical kinetic problems and estimating the molecular abundances \citep{gear71,hindmarsh95}.

 At each integration step, we must evaluate summations or differences of very small quantities, which involve a careful analysis. Summation of floating-point numbers is ubiquitous in numerical analysis and has been extensively studied \citep{higham93}. In \texttt{CHEMFAST,} we have rearranged the summation to calculate the differential abundances and minimize the numerical summation errors. Following this scheme, we can evaluate the very low abundances of chemical components.
We adopted the initial abundances of atoms and ions out of the SBBN theoretical predictions, based on recent determinations of $^4H_e$ abundances, as given in \cite{pitrou18}, namely,  $[D/H] \sim 2.46 \times 10^{-5}$, and the helium mass fraction, $Y_p$ = 0.24709. \\

\texttt{CHEMFAST} solves a three-level hydrogen atom model, in the same way as the \texttt{RECFAST}~\citep{seager99,seager00} recombination code, correcting the recombination rate to account for the superior excitation levels, but unlike the RECFAST code, we considered “coupling” with all molecular processes. This method has now been supplanted by state-of-the-art codes such as  \texttt{COSMOREC}~\citep{chluba10} and \texttt{HYREC}~\citep{alihaimoud11,lee20}, which developed a recombination computation using an effective multilevel atom method in order to meet the precision required for Planck \citep{Planck:2018vyg} CMB observation analysis.\\
Compared to the codes mentioned above, \texttt{CHEMFAST} additionally computes the chemical evolution of the homogeneous universe by following a whole chemical network. It proposes a simplified version compared to radiative transfer codes such as  \texttt{ENZO}~\citep{oshea04,bryan95,norman07}, which has also been used to follow primordial chemistry (see e.g.,~\citealt{schleicher08}).

\noindent
We start our calculations in a fully ionized Universe, at redshift $z_{\rm i}=10^4$. This redshift corresponds to the 
age of the Universe at $t_{\rm i}$, where $x=1+z$ :
\begin{equation}
t_{\rm i} \, = \, \frac{1}{H_0} \, \int_{1+z_{\rm i}}^\infty \, \frac{{\rm d}x}{x \, \sqrt{\Omega_{\rm r,0}x^4 + \Omega_{\rm m,0}x^3 + \Omega_{\Lambda,0}}} \ .
\label{time-z}
\end{equation}
\noindent

We stopped our calculations at a redshift of $z_{\rm f}=10$, without taking into account the reionization processes, as they are out of the scope of this work. \\

The initial number density of baryons $n_{\rm b,i}$ (i.e., number of baryons per cm$^{3}$ at $z_{\rm i}$) is computed as:
\begin{equation}
\overline{n}_{\rm b,i} \, = \, \frac{3 H_0^2 \, \Omega_{\rm b,i}}{ 8 \pi G\, \mu_{\rm b} m_{\rm p}} \, = \,
\frac{3 H_0^2 \, \Omega_{\rm b,0} }{ 8 \pi G \, \mu_{\rm b} m_{\rm p}} \, (1+ z_{\rm i} )^3 \ ,
\end{equation}
where $m_{\rm p}$ is the mass of a proton and $\mu_{\rm b}$ is the molecular weight per baryon particle. 
The molecular weight depends on the BBN initial conditions and is computed as \cite{seager99}:
\begin{equation}
\mu_b = \frac{1}{(1+f_{He})(1-Y_p)} ,
\end{equation}
where the helium number fraction, $f_{He}$, is given by
\begin{equation}
f_{He} = \frac{n_{He}}{n_H} \, = \, \frac{m_H}{m_{He}} \, \frac{Y_p}{1-Y_p}.
\end{equation}

Lastly, the initial radiation temperature is $T_\gamma(z_i) = T_{\gamma,0}(1+z_i)$. At the initial redshift we are considering here, the baryons and photons are fully thermally coupled, hence, we can take $T_k (z_i) = T_\gamma(z_i)$.

\subsection{Species abundances}

\label{sec:thermal}

The cosmological recombination process was not instantaneous because the electrons, captured into different atomic energy levels, could not cascade instantaneously down to the ground state.
The universe expanded and cooled faster than recombination could be completed and a small fraction of free electrons remained.
Recombination processes become dominant when the reactions of photoionization are negligible. We define the redshift of recombination, $z_{\rm rec}$, as the time at which the abundance of a neutral species is equal to the one of its corresponding ion. In this study, we followed an approximation of the multi-level atom model from~\cite{novosyadlyj22}, which reached the 400-level calculation result from the \texttt{Cloudy} code \cite{ferland17} with a $0.01\%$ precision. 
 In Fig. \ref{fig:atoms_exp}, we show the evolution of the relative abundances (i.e., the ratio of species number densities over the total baryon number  densities) of  all the atomic species considered in our reaction network. We find the successive redshifts of recombination of He$^{2+}$ ($z_{{\rm rec, He}^{2+}} \sim 5977$), He$^{+}$ ($z_{{\rm rec, He^+}} \sim 2556$), D$^{+}$ and H$^+$ ($z_{{\rm rec, D^+}} = z_{{\rm rec, H^+}} \sim 1387$). The flatness of the atomic relative abundances at  $z<100$ is caused by the inefficiency of collisional reactions due to the expansion of the Universe. It causes a decrease of species densities and freezes the relative abundances.

\begin{figure}[ht]
    \centering
    \includegraphics[width=0.49\textwidth]{./figures/figure1.pdf}
    \caption{Relative  number densities of atomic and ion species with respect to the total density of baryons, as a function of redshift. The computation goes from z=$10^4$ to $z=10$ Hydrogen species are represented as full lines, helium as dashed lines, and deuterium as dash-dotted lines. We indicate the recombination redshifts of primordial helium, deuterium and hydrogen nuclei (He$^{2+}$, He$^+$, then D$^+$ and H$^+$) with vertical dotted line.}
    \label{fig:atoms_exp}
\end{figure}

 \paragraph{\textbf{HeH$^+$ molecule}}
  Helium is the first neutral atom that appeared in the Universe (see Fig.~\ref{fig:atoms_exp}). One of the first molecular bonds to be formed is $\rm HeH^+$ via radiative association processes\footnote{$\rm He_2^+$ ions can form earlier but are quickly removed by photodissociation and dissociative recombination}: $\rm He^+ + H \rightarrow HeH+ + \nu$ (see \citealt{roberge82,kraemer95}) and $He +H^+ \rightarrow HeH^+ + \nu$ (see \citealt{zygelman98}). It then mostly contributes to the production of $H_2^+$ via exchange reaction ($\rm HeH^+ + H  \rightarrow H_2^+ + He$). ~\cite{schleicher08} also showed that this molecule can play a role in the smearing of primordial fluctuations by scattering CMB photons. Thus, helium is an important component to initiate, at the earliest times, the primordial chemistry.   Once a large number of helium atoms are formed through the recombination process, charge transfer with ions becomes possible, initiating the formation of the other neutral species. However, $\rm HeH^+$ appears to be negligible in the molecular cooling processes.

\begin{figure}[h]
    \centering
    \includegraphics[width= 0.49 \textwidth]{./figures/figure2.pdf}
    \caption{Relative number densities of molecular species with respect to the total density of baryons, as a function of redshift. The computation goes from $z_i=10^4$ to $z_f=10$.}
    \label{fig:molecules_exp}
\end{figure}
 
\paragraph{\textbf{Molecular hydrogen}} 
Molecular hydrogen cannot form directly by radiative association between two neutral atoms because H$_2$ does not have a permanent dipole moment. Charge transfer $ \rm H_2^+ + \rm  H \rightarrow \rm H_2 + \rm H^+$ becomes the only alternative.
Once HeH$^+$ was formed, this ion became an important source of H$_2^+$ production through the exchange reaction with some neutral H. Then, H$_2^+$ charge transfer with a neutral species led to H$_2$ molecules.
Moreover, as the radiation temperature decreases, H$_2$ can be formed through H$^-$ by radiative attachment $ \rm H + e^- \rightarrow \rm H^- + \gamma,$ followed by associative detachment $\rm H^- + \rm H \rightarrow \rm H_2 + e^-$~(\citealt{field00}). 
For densities over $10^8 \,\rm cm^{-3}$, $\rm H_2$ molecule can be effectively formed via three-body reaction $\rm H + \rm H + \rm H \rightarrow \rm H_2 + H$. However, the baryon density after thermal decoupling is $ \sim 550 \, \rm cm^{-3}$, which makes this channel ineffective in the context of a homogeneous Universe.
The two major routes of H$_2$ formation are visible in Fig.\ref{fig:molecules_exp} through the two jumps at the redshifts $z \sim 1000-500$ (H$_2^+$ channel) and $z\sim 150$ (H$^-$ channel). After the freeze-out of the relative abundances, we found $n_{\rm rel}(\rm H_2) = 2 \times 10^{-6 }$ for the $ \rm H_2$ molecule.
We notice that H$_3^+$ appeared in the early Universe. However, H$_3^+$ reactions, which are pivotal to the chemistry of dense interstellar clouds (see \citealt{oka80,tennyson95,herbst00}), do not play a significant role in primordial chemistry.

\paragraph{\textbf{Molecular deuterium}}
Deuterated-hydrogen molecules have permanent dipole moments, which provide them the capacity to be formed by radiative association (forbidden between two hydrogen atoms). Nevertheless, HD formation is significant when H$_2$ appeared, as the mechanism of  dissociative collision $ \rm H_2 + \rm D^+ \rightarrow \rm H^+ + \rm HD$ becomes efficient (see \citealt{palla95,stancil98a,signorepuy09}). Due to the slight difference between the electronic structure of D and H, H$_2$ and HD significantly appeared at the same epoch, as shown in Fig.\ref{fig:molecules_exp}.  The  relative abundance of the HD molecule lies at $n_{\rm rel}(\rm HD) = 6.1 \times 10^{-10}$  after the freeze-out.

\subsection{Molecular thermal functions}
The formation of primordial molecules such as $\rm H_2$ and $\rm HD$ generates thermal responses due to the excitation of rotational and vibrational levels of molecules. In the range of kinetic and radiation temperature considered, only the rotational levels can be excited.
Rotational level populations depend on  collisional reactions and radiative processes (CMB absorption or induced and spontaneous emission).
Two mechanisms are able to excite or de-excite molecules rotational levels, illustrated in Fig.~\ref{fig:rot_level}.
A molecule can be radiatively excited from a level $J$ to a superior one $J'$, by absorbing a photon external to the medium (for instance from the radiative background), then become de-excited via a collision with another atom or molecule from the medium. The external photon energy in converted into kinetic energy; hence,  we need to consider molecular heating. A molecule can be also excited by  a collision from  level $J$ to $J'$, then emit a photon by spontaneous or induced de-excitation to the $J$ level. If the photon is not reabsorbed by the medium, the associated energy is lost and we would be  considering molecular cooling.

\begin{figure}[ht]
\centering
\includegraphics[scale=0.22]{./figures/figure3.png}
\caption{Schematic illustration of the transitions between 2 rotational levels $J$ and $J'$. \textbf{Left:}  Molecule is excited to $J'$ level
by collision (double arrow), then de-excited by photon emission (wavy arrow). Assuming an optically thin medium, the photon is not reabsorbed, and the medium is cooled.  \textbf{Right:}   Molecule is excited by absorption of a photon (single arrow), then de-excited by collisions. The transition energy injected by the photon is then transmitted to the medium in the form of kinetic energy and the medium is heated.}
\label{fig:rot_level}
\end{figure}

To express the energy per volume unit that can be gained, $\Gamma_{\rm mol}$ (heating), or lost, $\Lambda_{\rm mol}$ (cooling), by the medium due to rotational level transitions, we define the thermal molecular function : 
\begin{equation}
    \Psi_{\rm mol} = \Gamma_{\rm mol} - \Lambda_{\rm mol}.
\end{equation}
We detail in Appendices~\ref{appendix:molthermalfunctions} and \ref{app:molpoplevels},  the computation of the heating and cooling terms, as well as the rotational level populations.

\subsubsection{Molecules thermal influence}

H$_2$ is homonuclear, which implies that it does not possess a permanent dipole moment, which strongly diminish its thermal effect. However, as it is the most abundant molecule, the thermal function due to H$_2$ plays a role in numerous astrophysical medium \citep{lebourlot99}. In the post-recombination context, H$_2$  contributes to heat the medium. Nevertheless, in Fig.~\ref{fig:cooling_exp}, where we compare the thermal channel contributions in $T_k$ evolution, we can see that the influence of H$_2$ on the background gas temperature (see Eq.~\ref{eq:tkin_exp}) is negligible (as already shown in \citealt{puy93} and \citealt{Pfenniger:2002ji}).\\

Infrared emissivities of HD were computed by ~\cite{dalgarno72}. The existence of a permanent dipole moment  and its excitation temperature (21K compared to 112K for H$_2$) give this molecule interesting cooling properties. ~\cite{puy97} showed, in a collapsing molecular protocloud,  HD is the main cooling agent around 200 K, confirmed by \cite{flower00b} and \cite{nakamura02}.
Despite being by far less abundant than  H$_2$, the HD molecule can still provide a considerable thermal contribution at the lowest temperatures, around $T_k<200K$ because of its permanent dipole \citep{puy93,Galli:1998dh,stancil98b}. However, as H$_2$, HD cooling does not play a role in the evolution of the gas temperature in a homogeneous expanding universe \citep{flower00a,Galli:1998dh,Galli:2002du}.
Thus,  the molecular thermal function, $\Psi_{\rm mol}$, is negligible in the thermochemistry of the homogeneous Universe. We will see its importance in a scenario of gravitational collapse in Sect. \ref{sect:collapse}.

\begin{figure}[ht]
    \centering
    \includegraphics[width=0.48\textwidth]{./figures/figure4.pdf}
    \caption{Evolution of cooling and heating components of the gas temperature as a function of redshift, in the homogeneous expanding universe scenario. All the components are described in terms of energy density per second. $\Psi_{\rm adia}$ arises from the expansion of the universe. As it is negative (cooling component), we display it as an absolute value. $\Psi_{\rm com}$ is due to Compton coupling between matter and radiation. $\Psi_{ \rm mol}=\Psi_{\rm H_2} +\Psi_{\rm HD}$ corresponds to the thermal contribution of molecules H$_2$ and HD.}
    \label{fig:cooling_exp}
\end{figure}

\section{21cm line implementation}
\label{sect:21cm}

The hyperfine transition line of atomic hydrogen, in the ground state, arises due to the interaction between the electron and proton spins. In the excited triplet state, the spins are parallel, whereas the spins in the ground (singlet) state are antiparallel. The 21cm line is a forbidden line for which the probability for a spontaneous $1 \longrightarrow 0$ transition is given by the Einstein coefficient $A_{10} = 2.85 \times 10^{-15} s^{-1}$.  21cm signal observable quantity is the differential brightness temperature,  which can be expressed as (see e.g.,~\citealt{madau97}) 
\begin{equation}
    \delta T_b \approx \frac{3 h c^3}{32 \pi}
\, \frac{A_{10}}{k_B \nu_{10}^2} \, \frac{\bar{n}_{HI} (1+\delta_b)}{(1+z)H(z)\left(1 + \frac{1+z}{H(z)}\frac{dv_{||}}{dr_{||}}\right)} \,\left( 1 - \frac{T_\gamma}{T_s}\right) , \\ 
\label{eq:dTb_ch4}
\end{equation}
where $\nu_ {10}  = 1420.406 \, \rm MHz$ is the 21cm line rest frequency, $\bar{n}_{\rm HI}$ is the mean neutral hydrogen number density, and  $\delta_b$ is the relative baryon density perturbation.  H(z) is the Hubble rate, and $dv_{||}/dr_{||}$ is the gradient of line of sight peculiar velocity, describing the relative motion of the gas with respect to the Hubble flow.
Lastly, $T_\gamma$ is the radio background temperature, which we set as the CMB one, and $ T_s $ is the spin temperature defined from the ratio between the hyperfine population levels.\\
 
Hydrogen atoms can be excited by various processes:  the absorption and stimulated emission of photons from the CMB redshifted at the 21cm wavelength, collisions with other hydrogen atoms, free electrons and protons, and  Lyman-$\alpha$ radiation from astrophysical sources through Wouthuysen-Field effect \citep{wouthuysen52,field58}. During the dark ages, before the formation of the first stars, only  the first two mechanisms compete in the excitation of the hyperfine level. In thermal equilibrium, the spin temperature can be written: 

\begin{equation}
T_s^{-1} = \frac{T_\gamma^{-1} + x_c T_k^{-1}} {1 + x_c}.
\label{eq:tspin}
\end{equation}
\noindent
$x_c$ is the total coupling coefficient for collisions given by the expression:
\begin{equation}
x_c = \frac{ \Bigl( C_H + C_e + C_p \Bigr)}{A_{10}} \,
\frac{T_{10}}{ T_\gamma},
\label{eq:xc_ch4}
\end{equation}

where $T_{10} = h\nu_{10}/k_b$ is the equivalent temperature of energy splitting between the two hyperfine levels. $C_H$, $C_e$, and $C_p$ are the de-excitation rates of the triplet due to collisions with neutral hydrogen, electrons, and protons. The collision term for each $i$ interacting species can be written as $C_i = n_i\, \kappa^i_{10}(T_k)$, where $\kappa^i_{10}$ is the effective single-atom rate coefficient for the transition 1-0 from collisions with that species.
These rate coefficients are obtained through quantum mechanics computation for collisions with other hydrogen atoms  \citep{zygelman05}, with free electrons \citep{furlanetto07a}, and with protons~\citep{furlanetto07b}. \\

In the context of homogeneous expanding universe, the brightness temperature of Eq (\ref{eq:dTb_ch4}) can be approximated as 
\begin{align}
    \delta T_b & \simeq \frac{3 h c^3}{32 \pi}
\, \frac{A_{10}}{k_B \nu_{10}^2} \, \frac{n_{HI}}{H(z) (1+z)} \, \left(1 - \frac{T_\gamma}{T_s}\right) ,\\
& \simeq   27 x_{HI} \left(\frac{\Omega_b h^2}{0.023} \right) \sqrt{\frac{0.15}{\Omega_m h^2} \frac{1+z}{10}}  \left( 1 - \frac{T_{\gamma}}{T_s}\right).
\label{eq:dTb_approx_ch4}
\end{align}
As the universe is considered homogeneous, we only keep the mean hydrogen density $\bar{n}_{HI}$, and set $\delta_b=0$.  Moreover, we consider an average peculiar velocity equal to zero.
The second expression explicits the dependence of the 21cm signal to cosmological parameters $\Omega_b$  and $\Omega_m$, the baryon and matter energy density fraction today, and $h$, the Hubble parameter in units of 100 km/s/Mpc. Lastly, $x_{HI}$  is the hydrogen neutral fraction. We show in Fig~\ref{fig:tktrts_tb_exp} (top), the evolution of spin, radiation and kinetic temperatures. The behavior of the spin temperature can be separated in three distinct epochs that determine the observability of the brightness temperature, which is visible in the bottom panel of Fig.~\ref{fig:tktrts_tb_exp}. These epochs are described as follows.  
\begin{itemize}
    \item Before matter-radiation decoupling, the spin temperature is coupled to the CMB temperature ($T_s = T_\gamma$). Differential brightness temperature, $\delta T_b$, remains at zero.
    \vspace{0.2cm}
    \item Around $z\sim z_{dec,1\%}$ (defined as the redshift at which $T_k(z_{\rm dec,1\%}) = 99\% T_\gamma(z_{\rm dec,1\%})$), matter and radiation temperatures are progressively decoupling. The Universe is dense enough for the collision mechanism to dominate in the 21cm signal production. The spin temperature is thermalized  to the gas temperature, $T_s \rightarrow T_k$, and the brightness temperature reaches a minimum of $ \delta T_b\sim -44$ mK at $z(\delta T_{b,min} \approx 89$).
    \vspace{0.2cm}
    \item Due to the expansion of the universe, the baryons get more and more diluted and the collision mechanism progressively becomes ineffective. The spin temperature relaxes to the CMB temperature, $T_s \rightarrow T_\gamma$, which brings $\delta T_b$  back to zero.
    \end{itemize}
    
\begin{figure}[ht]
    \centering
    \includegraphics[width=0.9\linewidth]{./figures/figure5.pdf}
    \caption{\textbf{Top}: Evolution of kinetic gas temperature, $T_k$ (blue solid line), radiation temperature, $T_\gamma$ (dashed green line), and 21cm spin temperature, $T_s$ (orange dash-dotted line), as a function of redshift. \textbf{Bottom}: Evolution of the  global 21cm differential brightness temperature as a function of redshift. At $z_{\rm dec,1\%}$ ($T_k = 99\% \, T_\gamma$), thermal decoupling starts to be effective. At $z(\delta T_{b,\rm min})$, $\delta T_b$ reaches its lowest point, corresponding to the maximum absorption during the dark ages. }
    \label{fig:tktrts_tb_exp}
\end{figure}

This section has served to present the functioning of \texttt{CHEMFAST} within the known framework of an expanding homogeneous universe. The code can be used to solve a system of stiff coupled differential equations in order to obtain the evolution of the abundance of chemical species based on a network of reactions.
The code can also be used to calculate the thermal influence of molecules on temperature through the excitation of their rotational levels. In a homogeneous expanding universe, we can see that this effect is negligible.
Finally, by following the thermal history of the universe and the chemical reactions, we can estimate the intensity of the global 21cm signal due to arising from collisional coupling. The 21cm global signal computed with \texttt{CHEMFAST} is in accordance with the state-of-the-art literature (see e.g., \citealt{furlanetto06,Pritchard:2011xb}).

\section{Dynamics of collapsing clouds and 21cm hydrogen line emission}

\label{sect:collapse}

In the context of homogeneous expansion seen in the first part, the thermal influence of $H_2$ and $HD$ molecules on the gas temperature is negligible, due to their very low mean abundances. Their contribution is completely drowned out by the adiabatic cooling of the matter-radiation thermal coupling associated with Compton scattering. However, we know that the universe is actually not homogeneous, but exhibited small primordial density fluctuations,  which became seeds for the formation of structures.  \\
  
When a density perturbation grows enough due to gravity in order to reach a density contrast comparable to unity, it can no longer be described by the linear theory of perturbations. We must then consider a model to follow its evolution until a potential collapse. Molecules in overdensities are known to cool the gas via the excitation of their rotational and vibrational levels and permitting the further gravitational collapse of the gas, thereby setting up the right conditions for the formation of the first structures.

In this section we focus on the evolution of an individual overdensity.  We implement a collapse model in \texttt{CHEMFAST} by modifying the equations of dynamics. Our goal is then to estimate the 21cm signal from the collapsing overdensity under the influence of molecular cooling.

\subsection{Model of collapsing cloud}

We consider a spherical pressure-supported collapse model with basic hypothesis. The model we use is based on the work of~\cite{lahav86}, extended by~\cite{puy96}. Unlike the free-fall  model, the spherical collapse we consider cannot be solved analytically and requires a numerical resolution.

We assume that the overdense region we follow is isothermal, spherical, and without rotation. We also assume the  matter density to be constant inside the overdense region. This matter sphere, at first following the expansion of the universe, will progressively slow down with respect to the background, due to the density excess, until the point when it drops out of expansion, and starts to collapse back on itself. In the range of redshift we consider, the universe is fully dominated by matter, and we can assume a linear $t^{2/3}$ growth  of the fluctuations with the expansion of the Universe. Hence, we consider a spectrum of isothermal perturbations, that are described by the spectrum \citep{gott75}: 
\begin{equation}
\frac{\delta \rho}{\bar{\rho}}  = \left(\frac{M}{M_0}\right)^{-1/3}(1+z)^{-1} ,
\label{dens_spectrum}
\end{equation}
where the mass $M$ of the overdense region is normalized to $M_0$, defined as a characteristic mass-scale of  $M_0=10^{15} M_{\odot}$, which is the typical mass of large galaxy clusters today, also used in \cite{lahav86}.

\subsection{Equations}
As in Sect. \ref{sec:homogeneous}, here we solve a set of differential equations, this time for a collapsing overdense region.
\paragraph{\textbf{Baryon number density}}

The matter conservation inside the cloud gives : 
\begin{equation}
    \frac{dn_b}{dt}=  -3 \frac{n_b}{r} v_r + \sum_\xi \left(\frac{dn_\xi}{dt}\right)_{\rm chem}, \ {\rm{with}} \ 
v_r =\frac{dr}{dt}.   
\end{equation}
The equation is similar to the expansion case of Sect. \ref{sec:homogeneous}, the scale factor, $a$, is simply replaced by the radius of the cloud, $r$.  The last term in the right-hand side of the equation is the contribution from chemical reactions. 

\paragraph{\textbf{Species densities}}

Similarly, for each chemical component $\xi$ we have
\begin{equation}
\frac{d n_\xi}{ dt} = - 3  \, \frac{n_\xi}{r}v_r
+ \Bigr( \frac{{\rm d}n_\xi}{{\rm d}t} \Bigl)_{\rm chem}
\label{xi_eff}
.\end{equation}

\paragraph{\textbf{Temperatures}}
The radiation temperature evolves just like in the homogeneous case, with the expansion of the universe:
\begin{equation}
    \frac{dT_\gamma}{dt} = - H(t) \, T_\gamma.
\end{equation}

The kinetic gas temperature, under the assumption of perfect gas, can be computed as:
 \begin{equation}
    \frac{dT_k}{dt} \, = \,  
    \frac{2}{3n_b k_B}\left(\Psi_{\rm grav } + \Psi_{\rm mol} +\Psi_{\rm com}\right) .
    \label{eq:tk_eff}
    \end{equation}
  $\Psi_{\rm grav}= -3 n_b k_B T_k \frac{v_r}{r} $ accounts for the density variation arising from the gravitational collapse,  $\Psi_{\rm com}$ describes the Compton coupling between matter and radiation (see Eq.~\ref{eq:compton}), and  $\Psi_{\rm mol}$ expresses the thermal influence of molecules on the gas temperature through the excitation of their rotational levels (see Appendix \ref{appendix:molthermalfunctions}).

\paragraph{\textbf{Radius and velocity of collapse}}

Starting from the energy balance: 
\begin{equation}
     E = -\frac{3}{5} \frac{GM^2}{r} + \frac{3}{10}M v_r ^2 + \frac{3}{2}N_b k_b T_k,
\end{equation}
where the first term corresponds to the potential energy in a spherical system with a central potential. The second one accounts for the associated kinetic energy, and the last one for the internal energy. Then, $N_b$ the total number of baryons in the cloud.\\
We inject the temperature into Eq.~\ref{eq:tk_eff}, expressing the energy variation in the system as  $ \frac{dE}{dt} = r^3(\Psi_{\rm com} + \Psi_{\rm mol})$, and use M = $N_b \, \mu_b m_H \frac{\Omega_m}{\Omega_b}$\footnote{We suppose the balance between baryons and dark matter is identical as the background.} to obtain 
   
\begin{equation}
    a_r = -\frac{GM}{r^2} + \frac{5k_bT_k}{\mu m_H} \, \frac{1}{r} \frac{\Omega_b}{\Omega_m} 
    \  \ {\rm{with}} \  \
    a_r = \frac{dv_r}{dt}.
     \label{ar}
\end{equation}

\subsection{Initial conditions}
 Considering a single cloud of mass, $M$, we follow the collapse starting from the turnaround point, when the gravity takes over the expansion and the perturbation starts to increase in density, as it was only diluting slower than the background before this moment.
In a spherical collapse model, an overdense region reaches the turnaround point when the density contrast reaches  $\Bigl(\frac{3\pi}{4}\Bigr)^2$ \citep{padmanabhan02,peebles80}. 

Hence, the baryon number density at turnaround is expressed by  
\begin{equation}
    n_{b,ta} =\Bigl(\frac{3\pi}{4}\Bigr)^2  \, \bar{n}_b(z_{ta}) \, .
\end{equation}
Where $\bar{n}_b(z_{ta})$ is the background baryon number density  at turn-around redshift, $z_{ta}$.\\
Following the approach of~\cite{gunn72} and ~\cite{lahav86}, we compute the redshift of turn-around for a given mass:
\begin{equation}
    1+ z_{ta} = \left( \frac{3 \pi}{4}\right)^{-2/3} \, \left( \frac{M}{M_0}\right)^{-1/3}
.\end{equation}
This is the redshift at which we start to follow the collapse in \texttt{CHEMFAST}. It is only dependent of the mass, $M,$ of the cloud.\\

The radius of the overdense region is directly derived from the density 
\begin{equation}
    r_{ta} = \left(\frac{3M}{4\pi \rho_{b,ta}}\right)^{1/3} \, ,
\end{equation}
with $\rho_{b,ta} = \mu_b m_H \, n_{b,ta}$.
The collapse velocity $v_{ta}$ at turn-around is by definition set at zero: 
\begin{equation}
v_{ta} \, = \, \frac{dr_{ta}}{dt} = 0.
\end{equation}

For the gas temperature, the computation is less evident. If the gas and radiation temperature were fully decoupled, we could consider the adiabatic gas temperature evolves as $T_k(z) \propto \rho_b(z)^{2/3}$. Then, the gas temperature inside an overdensity at turnaround  redshift could be written 
\begin{equation}
    T_{k,ta}^{\rm adia} = \left(\frac{3\pi}{4}\right)^{4/3} \bar{T}_k (z_{ta}) \, ,
\end{equation}
where the $ \rm adia$ label refers to adiabatic, in the sense that the gas temperature only changes due tot the expansion or compression of its environment, without any exchange of heat. However, as we will see, our model can provide very high redshift of turn-around, depending on the chosen mass. As a consequence, the gas temperature does not fully evolve adiabatically, and the Compton scattering between free electrons and photons can still be efficient at the turn-around redshifts. Choosing $T_{k,ta}^{\rm adia}$ would mean ignoring the effect of Compton scattering in the period preceding the turnaround. 
We therefore need to estimate the temperature at turn-around, which lies in the range $\left[\overline{T}_k (z_{ta}),  \left(\frac{3\pi}{4}\right)^{4/3} \overline{T}_k (z_{ta})\right]$, accounting for the residual Compton coupling. 
We chose the turnaround temperature empirically, such that the Compton thermal channel, $\Psi_{\rm com}$, is steady in the very first moments of collapse at the beginning of the run. Actually, we have checked that this detail does not significantly impact the behavior of $T_{k}$ during the rest of the collapse. It does, however, help to avoid non-physical behavior\footnote{In particular, we avoid having a very high $T_k$ at turn-around, which decreases drastically when the run starts, as if we just added a Compton coupling that was not here before turnaround.} in the very first moments of the collapse.\\

To summarize, our model takes the mass of the overdense region as the only free parameter, from which all the initial conditions are derived.
For a given mass, we ran  \texttt{CHEMFAST} a first time in expansion mode up to the turnaround redshift and extracted the background values of the species abundances, baryon density, and temperatures. These data were then used to compute the initial conditions in the collapse setup. \\

\section{Results}

\label{sect:results}

\subsection{Thermal evolution of $10^8 \rm M_{\odot}$  collapsing cloud}
We focus on the analysis of  a $10^8 M_{\odot}$ collapsing overdensity, which corresponds to the order of magnitude expected for "massive minihalos," the least massive halos that are expected to be able to host star formation (see e.g., \cite{Tegmark:1996yt, yoshida03, glover05, Haiman:2006si, Trenti:2009cj, greif15} for discussions on the formation of the first structures). In our model, this overdensity reaches its redshift of turnaround at $z_{ta} = 93$, with a temperature $T_{k,ta} = 2.1\cdot \overline{T}_{k}(z_{ta})$. We first discuss the evolution of the gas temperature from its different thermal channels and then propose an estimate of the 21cm line arising from the collapsing overdensity.
\vskip2mm
\noindent

For all the following figures, we display the  evolution of the quantities as a function of the time to collapse. Defining the turnaround time: 
\begin{equation}
    t_{ta} = \sqrt{\frac{3\pi}{32G\rho_{ta}}} ,
\end{equation}
the time of re-collapse can be expressed as $t_{\rm coll} = 2 t_{ta}$ \citep{lahav86}. We normalized time in the figures to evolve between 0 and 1.\\

In Fig.~\ref{fig:cooling_eff8}, we follow the intensity of all the thermal channels acting on the gas temperature within the cloud (see Eq.~\ref{eq:tk_eff}).
The heating due to gravitational contraction, $\Psi_{\rm grav}$ (solid blue line), gets stronger as the radius of the halo decreases and the collapse velocity increases; $\Psi_{\rm com}$ (dashed curve)  arises from the Compton coupling between matter and radiation. Since our initial conditions provide a gas temperature in the halo above the background radiation temperature, the Compton interaction tends to cool the gas temperature towards the radiation temperature, which explains why it presents a negative contribution. The  Compton coupling dominates over the other contributions in the first instants and decreases as the collapse progresses. It eventually becomes negligible. \\

The $\Psi_{\rm mol}$ term (dotted curve) represents the molecular thermal function summing over the contributions from the two most important molecules $\Psi_{\rm H_2}$ and $\Psi_{\rm HD}$. Furthermore, $\Psi_{\rm mol}$ is almost completely driven by $ \rm H_2$ molecule. The HD contribution becomes important for lower temperatures than the one we are interested in (< 200 K), it is negligible in the case we are studying~\citep{mcgreer08}.
In Fig.~\ref{fig:cooling_eff8}, $\Psi_{\rm mol}$ contributes to cool the gas. Initially negligible, its importance increases sharply as the density of the molecules increases and the gas temperature grows, until it becomes significant and even dominates the other thermal channels at the end of the collapse. Compared to the expansion scenario, the molecular thermal function has a strong influence on the gas temperature.
\vskip2mm

\begin{figure}[ht]
    \centering
    \includegraphics[width=0.48\textwidth]{./figures/figure6.pdf}
    \caption{Evolution of cooling and heating components of the gas temperature in a $10^8 M_{\odot}$ collapsing cloud as a function of time, normalized by $t_{\rm coll}$.  All the components are described in terms of energy density per second.
    $\Psi_{\rm grav}$ arises from to gravitational collapse of the gas;
    $\Psi_{\rm com}$ is due to Compton coupling. Finally,
    $\Psi_{ \rm mol}=\Psi_{\rm H_2} +\Psi_{\rm HD}$ corresponds to the thermal contribution of molecules H$_2$ and HD.
    The negative components  $\Psi_{\rm com}$, $\Psi_{\rm mol}$, $\Psi_{\rm H_2} $,  and $\Psi_{\rm HD}$ are displayed as absolute values.} 
    
    \label{fig:cooling_eff8}
\end{figure}

We display the evolution of the radiation and gas temperatures inside the $10^8 M_\odot$ collapsing overdensity  in Fig.\ref{fig:tktrcoll} (top panel). The radiation temperature evolves exactly in the same way as in the background, and is shown for comparison. The evolution of the gas temperature can be divided in several stages: 
\begin{itemize}
\item During  the very first moments of the collapse, Compton coupling is the dominant thermal mechanism and tends to bring $T_k$ (blue solid line) to $T_\gamma$ (green dashed line) acting against gravitational contraction. As a result, the gas temperature inside the overdensity slightly decreases. This stage holds for a very short time; hence, the decrease is not visible in the figure.

\item Heating due to the gravitational collapse rapidly dominates over the Compton coupling, starting from $t = 0.06 t_{\rm coll}$. From this moment, the gas temperature inside the cloud grows drastically, under the influence of gravity.

\item Molecular cooling rapidly becomes more important due to the increasing gas temperature and H$_2$ density inside the cloud. This thermal channel dominates, starting from $t = 0.89 t_{\rm coll}$, the molecular thermal function cools the medium more than it heats up by gravitational contraction. The gas temperature inside the cloud decreases despite the ongoing collapse.

\end{itemize}

\begin{figure}[ht]
    \centering
    \includegraphics[width=0.48\textwidth]{./figures/figure7.pdf}
    \caption{\textbf{Top:} Evolution of radiation temperature, $T_\gamma$ (dash-dotted line), gas temperature, $T_k$ (solid line) and 21cm spin temperature, $T_s$ (dotted line) for a collapsing $10^8 M_{\odot}$ cloud, as a function of the time to collapse. We highlight on top of the figure the times when the dominant thermal channel changes. \textbf{Bottom:} Evolution of the associated 21cm brightness temperature from a collapsing $10^8 M_{\odot}$ cloud.}
    \label{fig:tktrcoll}
\end{figure}

\subsection{21cm line emission from collapsing halo}

At the top of Fig.~\ref{fig:tktrcoll}, we can see the spin temperature, $T_s$ (dashed line), closely follows gas temperature, $T_k$, throughout the collapse. Looking again at the spin temperature (Eq.~\ref{eq:tspin}), we can explain it by the following arguments. We are in a collapse scenario, where the matter number density is increased up to a factor $\sim 177$  compared to the background, at the end of the collapse. Therefore, the collisional coupling is important and dominates the 21cm line production. 

The differential brightness temperature, $ \delta T_b$, is positive, meaning that the signal presents an emission feature, different from the global signal absorption feature from the expansion scenario.
Throughout the collapse,  $T_s$ is always greater than radiation temperature, $T_\gamma$. Therefore, the term $\left(1 - \frac{T_\gamma}{T_s}\right)$ in Eq.~\ref{eq:dTb_approx_ch4}  stays between 0 and 1.
 Moreover, we observe in $ \delta T_b$ the same type behavior than $T_k$, with a turnover caused by molecular cooling. Indeed, since the spin temperature completely follows $T_k$, it is sensitive to the same thermal processes, transmitted to $ \delta T_b$.
We can note that, even without the activation of molecular cooling, $ \delta T_b$ would eventually saturate and reach a maximum limit when $T_s \gg T_\gamma$. In this regime, the dependence on $T_s$ in Eq.~\ref{eq:dTb_approx_ch4} would become unimportant.\\

We adopted the Eq.~\ref{eq:dTb_approx_ch4} for the computation of the 21cm differential brightness temperature. By doing so, we made several approximations, to which we dedicate the following discussion. Firstly, we assumed a zero peculiar velocity for the halo. This choice is not problematic by itself, but it should be noted that it is arbitrary. When studying a  whole population of halos, we have to take into account the effects of peculiar velocities, which induce an enhancement of the fluctuations at all scales \citep{bharadwaj04,barkana05}.\\

Equation~\ref{eq:dTb_approx_ch4} also implies two major approximations for calculating the 21cm signal from a halo. The supposition that the optical depth is thin  can no longer technically hold  in a halo. For an accurate calculation, it would be more appropriate to consider a density profile (such as a truncated isothermal sphere used in \citealt{iliev02}, or a Moore profile used in \citealt{furugori20}) for the halo; instead of a homogeneous density as we have done. The optical depth would then vary as a function of radius (maximum at the center of the halo where the maximum amount of matter is traversed)  and the brightness temperature shall be integrated over the surface of the halo. \\
The other approximation we make is to assume the 21cm line profile as a Dirac function. A more appropriate approximation, such as a thermal-Doppler-broadened line profile, would allow the velocity dispersion of particles within the halo to be taken into account.\\

However, accounting for the variable optical depth and the Doppler broadening would require us to get rid of the assumption of homogeneous density and temperature in the halo;  it could also strongly modify the 21cm signal. The results we show are in that sense approximations, presenting our approach, and we leave the improvement of the collapse model to a future work.

\subsection{Thermal evolution of a range of halo masses}
After focusing on the physics at work in a $10^8 M_\odot$ collapsing halo, we present the temperatures evolution for a whole range of different masses in Fig.~\ref{fig:mass_range} and display their associated turnaround redshifts and gas temperature in Table~\ref{tab:params_ta}. 
 Several masses are shown in the same figure as a function of their time to collapse in order to easily compare them. However, it is important to bear in mind that these collapses take place at different redshifts, which are correlated with the initial mass.\\

\begin{figure}[ht]
     \centering
     \begin{subfigure}
         \centering
         \includegraphics[width=0.49\textwidth]{./figures/figure8a.pdf}
     \end{subfigure}
     \begin{subfigure}
         \centering
         \includegraphics[width=0.49\textwidth]{./figures/figure8b.pdf}
     \end{subfigure}
        \caption{\textbf{Left:} Evolution of  gas temperature, $T_k$ (solid lines), for clouds of mass between $10^{6.0}$ and $10^{8.0}$ $M_\odot$, as a function of the time to collapse (top). On the bottom of this plot are displayed the  21cm differential brightness temperatures. \textbf{Right:} Same as the left panel, for masses between $10^{8.0}$ and $10^{8.8}$ $M_\odot$, also displaying the 21cm spin temperature $T_s$ (dotted lines). }
        \label{fig:mass_range}
\end{figure}

 In the left panel of Fig.~\ref{fig:mass_range}, we show the temperatures evolution for masses between $10^6$ and $10^8$ $M_\odot$. The gas temperature presents the same general behavior for all the masses: a growth provoked by the contraction of the clouds, followed by a decline in the last moments of the collapse, due to the molecular cooling. However, the fall in $T_k$ occurs at different moments of the collapse depending on the mass. It happens later for the smaller masses, such as $10^6 M_\odot$,  resulting in a higher maximal gas temperature, and 21cm brightness temperature.\\

On the other hand, we notice an inversion of this behavior starting from masses~$>10^{7.0} M_\odot$. We observe that the maximum gas temperature increases again and the moment of collapse at which the maximum temperature is reached is delayed.\\

Looking at the left of  Fig.~\ref{fig:cooling_mass_range}, where is plotted the ratio between $\Psi_{\rm mol}$ and $\Psi_{\rm grav}$, we observe that above $ 10^{7} M_\odot$,  the more we increase the mass of the cloud, the later the molecular cooling dominates. This effect can be understood by looking at the number density of H$_2$ inside the halos, presented in the right panel of Fig.~\ref{fig:cooling_mass_range}. We can expect a higher overall baryon density in the halos of smaller mass, as they are the first ones to form. The density contrast between the overdense region and the background is similar for all halos, so the one forming when the background is denser are also expected to show a higher density. However, the H$_2$ density directly depends on the presence of the species leading to its formation, H$_2^+$ via charge transfer reaction $\rm H_2^+ + H \rightarrow \rm H_2 + H$, and H$^-$ via associative detachment $\rm H^- + H \rightarrow \rm H_2 + e^-$. Finally,  H$^-$ is actually rapidly destroyed by CMB photons when forming at very high redshift, before it would be able to contribute to H$_2$ formation. \\

\begin{figure}[ht]
     \centering
     \begin{subfigure}
         \centering
         \includegraphics[width=0.48\textwidth]{./figures/figure9a.pdf}
     \end{subfigure}   
     \begin{subfigure}
         \centering
         \includegraphics[width=0.48\textwidth]{./figures/figure9b.pdf}
     \end{subfigure}
        \caption{\textbf{Left:} Evolution of ratio between $\Psi_{\rm mol}$ and $\Psi_{\rm grav}$ thermal channels, arising from  molecular excitation  and gravitational contraction respectively, for clouds between $10^{6.0}$ and $10^{8.0}$ $M_\odot$, as a function of the time to collapse. \textbf{Right:} $H_2$ number density evolution as a function of the time to collapse, for clouds between $10^{6.0}$ and $10^{8.0}$ $M_\odot$. }
        \label{fig:cooling_mass_range}
\end{figure}

For the lowest masses (e.g., $10^6 M_\odot$, which has a turn-around redshift of $z_{ta}=436$ in our model), despite a high overall density in the cloud, the formation of H$_2$ molecules by the H$^-$ channel is strongly suppressed. Up to $\sim 10^7 M_\odot$ ( $z_{ta} =202$), the H$^-$ are progressively less destroyed, allowing more H$_2$ to be formed and resulting in stronger molecular cooling. 
 In the homogeneous universe (see Fig. \ref{fig:molecules_exp}), the growth of the H$_2$ population via H$^-$ takes place around $z=150$, when the temperature of the photons decreases enough such that it does not allow them to destroy H$^-$ anymore, letting it maximally convert into H$_2$.
However, as the $> 10^7 M_\odot$ halos form later and later, even if the H$^-$ population is not suppressed, the baryons density (and therefore the H$_2$ density) progressively decreases, explaining the gradual decline of molecular cooling. \\

Interestingly, the 21cm brightness temperature does not present the same behavior and the highest maximum $\delta T_b$ is still reached in the lightest halos. The reason lies in the redshift dependence of the signal, which is not visible in the figure, but is explicit in  Eq.~\ref{eq:dTb_approx_ch4}. The collapses of the largest masses occur too late and the gas temperature increase is not enough to overcome the decline in redshift, even though $T_s$ is still strongly coupled to $T_k$ via collisions, as can be seen in the right panel of Fig.~\ref{fig:mass_range}.

\begin{table*}[ht]
    \caption{Redshifts and temperatures of turnaround}
    \centering
    \scalebox{1.15}{
    \begin{tabular}{ |c|c c c c c c c c|}
    \hline
        $\log_{10}$M [$M_\odot$] & 6.0             & 6.5               & 7.0               & 7.5                 & 8.0              & 8.2 & 8.5 & 8.8                 \\
        \hline
         $z_{ta}$ & 436   & 297      &  202   &137  & 93   &  80 &63& 50\\  
         $z_{\rm coll}$ &185 & 127 & 86 & 59 & 40 &34 & 26 & 21\\
         $T_{k,ta}$ (factor of $\overline{T}_k(z_{ta})$)& 1.05 & 1.14 & 1.30 &1.60 & 2.10& 2.40&2.90 & 3.14 \\
         \hline
    \end{tabular}}
    \vspace{0.5em}\\
     \noindent\textbf{Notes.} Temperature is shown as a factor of $\overline{T}_k(z_{ta})$, the background temperature at $z_{ta}$. $z_{\rm coll}$ is the redshift at which we stop the run, when the time to collapse $t_{\rm coll}$ is reached.
    \label{tab:params_ta}
\end{table*}

\section{Discussion and perspectives}

\label{sect:conclusion}
We  explored  the evolution of species abundances in detail during and after successive recombinations. To do so, we improved the code \texttt{CHEMFAST}, which solves a system of stiff coupled differential equations. Part of the system describes the dynamics (density, baryon, and radiation temperatures) and another is dedicated to describing the network of collisional, electronic, and photo-processes between the species. Particular attention has been paid to tracking the molecular abundances.\\

We first applied \texttt{CHEMFAST} to the simple expanding homogeneous universe framework. We computed the successive recombinations of $\rm He^{2+}$, $\rm He^{+}$, $\rm H^+$, and $\rm D^+$, as well as the relative abundances of all the atoms, ions, and molecules contributing to the reaction network. We also computed the thermal influence of molecules through the excitation of their rotational levels. Furthermore, we implemented the 21cm differential brightness temperature global signal computation during the dark ages, arising from collisional excitation. It presents an absorption peak with an intensity of -40 mK at z=89. This result is largely corroborated by the literature (e.g.,~\citealt{furlanetto06,Pritchard:2011xb}) and validates the effectiveness of our code\\

In a second study, our goal was to investigate the 21cm signal evolution in a new context and to assess how it could be distinguished from the sky-averaged global signal. To this end, we flexibly integrated a new set of dynamical equations into our abundance calculation code \texttt{CHEMFAST}, to follow the collapse of an overdense region. We based our work on a spherical collapse model developed by~\cite{lahav86}, taking into account the pressure forces,  with the cloud mass as the only free parameter. We followed the collapse of a typical cloud of $10^8 M_{\odot}$, expected to be able to host the formation of the first galaxies. Within the collapse, molecules are present in much greater density than in the homogeneous case.  Calculating the excitation of the rotational levels for the H$_2$ and HD molecules shows us that they have a strong thermal impact on the cloud's gas temperature. At this point, the contribution becomes dominant and takes over the gravitational heating in the last moments of the collapse, leading to a cooling of the gas. \\

Finally, we computed the  21cm line brightness temperature within the collapsing cloud. The signal shows a very different signature compared to the global signal of the dark ages. It presents an emission feature and is also affected by molecular cooling, in the same way as the gas temperature to which the brightness temperature is coupled via collisions. We also present the thermal evolution for overdense regions of $10^6$ to $10^{8.8}$ M$_\odot$. The strongest 21cm signals are expected from the halos of the lowest masses, which have lower H$_2$ abundances, due to the suppression of the H$^-$ formation channel by CMB radiation. These halos therefore exhibit less efficient molecular cooling. The intensity of molecular cooling increases progressively with the halo masses until redshifts $z\sim 150$ at which  H$^-$ is no longer destroyed. Halos whose formation begins after this redshift see a decline in molecular cooling, due to their lower density. Despite a higher $T_k$, this pattern is not followed by the 21cm brightness temperature because of its redshift dependence. The H$_2$ chemistry is not trivial, however, and further investigation would be required to understand  the  H$_2$ density evolution in collapse context, depending on the initial mass. We plan to conduct this investigation in a future work.\\

These particular signatures of the 21cm signal are promising with respect to probing the thermal history inside halos. They could also present an observational interest at the smallest scales of the 21cm power spectrum in the context of the forthcoming lunar-based observations of the dark ages.
One way to estimate this impact would be to compute the brightness temperature arising from a whole halo population, by associating our single halo results to the computation of a halo mass function~\citep{lukic07,murray13}. However, the particularly high turnaround redshifts of our collapse model make the determination of the halo mass function particularly challenging.\\ 
 
Indeed, our model predicts cloud formation at particularly high redshifts, compared to what is currently proposed in the literature (e.g., \citealt{Trenti:2009cj,Glover:2012gx,greif15}).  State-of-the-art halo formation models start at redshifts of $z\sim 30$ for the halos of the lowest mass, which are still allow the gas to collapse ($\sim 10^5 - 10^6 M_{\odot}$). Similar ranges are found by studies of 21cm signal predictions from halos \cite{iliev02,furugori20,Novosyadlyj24}, which makes it difficult to make comparisons with them, as our hypotheses are very different.
This discrepancy can be explained by the simplicity of the model we have chosen. Furthermore, we have seen that the assumption of homogeneous density and temperature inside the cloud could prove problematic for the study of the 21cm signal. We intend to study other collapse models with $\texttt{CHEMFAST}$ in future work and include the density and temperature profiles in our modeling. We also plan to integrate elliptical geometries, which are more consistent with collapse simulations. \\

Lastly, molecular cooling can induce the development of thermal instabilities. These effects were first studied by~\cite{parker53}, followed by ~\cite{field65}, and later ~\cite{balbus86}, who derived instability criteria, leading to a growth of the perturbations in the gas. It is now understood that these thermal instabilities play an important role in the fragmentation of halos and the formation of the first stars \citep{safranek10,Yoshida:2006bz,Bromm:2013iya,Glover:2012gx,greif13}.
 The aim of our simple model has been to highlight the crucial thermal influence of molecules on the thermal history and 21cm signal, but it does not allow us to explore the evolution of thermal instabilities in depth. We would then have to turn to adaptive mesh refining (AMR) type simulations, such as the \texttt{ENZO}~\citep{bryan14} or \texttt{RAMSES}~\citep{teyssier02} codes. This type of approach makes it possible to spatially monitor instabilities that can lead to fragmentation within the structures \citep{tang24}. 

\vspace{1cm}
\begin{acknowledgements}
We thank Alice Faure for her support in the code optimization. We also thank Daniel Pfenniger and Benoît Sémelin for helpful discussions.
These results have been made possible thanks to LUPM’s cloud computing infrastructure, founded by PHONE project - Occitanie Regions, Ocevu labex, and France-Grilles.
\end{acknowledgements}

\bibliographystyle{aa}
\bibliography{references}

\clearpage

\onecolumn
\begin{appendix}

\section{Details on molecular thermal function}
\label{appendix:molthermalfunctions}

In this appendix, we detail the heating and cooling terms composing the molecular thermal function: 
\textbf{\begin{equation}
    \Psi_{\rm mol} = \Gamma_{\rm mol} - \Lambda_{\rm mol}.
    \label{eq:psi_mol}
\end{equation}}

\subsection{Molecular heating}
The collisional de-excitation probability expresses the ratio between the collisional de-excitation rate and the total de-excitation rate, which includes collisions, as well as induced and spontaneous emissions, as follows: 
\begin{equation}
    P^c_{J',J} = \frac{n_\xi X_{J'} k^\xi_{J',J} } {n_\xi X_{J'}k^\xi_{J',J} + X_{J'}A_{J',J} + X_{J'}B_{J',J}\rho_{J',J}} \ .
\end{equation}
Here, $n_\xi$ is the numerical density of the collision partner species, $\xi$, and $X_{J'}$ is the population fraction of molecule in the rotational level $J'$. Then, $k^{\xi}_{J',J}$ is the collision rate of this reaction, while $A_{J',J}$ and $B_{J',J}$ are the Einstein coefficients for spontaneous and induced emission, respectively, and $\rho_{J',J}$ is the radiation energy density. Each term of the sum is the de-excitation term associated to one of the three mentioned processes.

Taking the product between the molecule density at $J$ level ,$n_{\rm mol}X_J$, the induced radiative excitation probability, $B_{J,J'}\rho_{J,J'}$, the collisional de-excitation probability, $P^c_{J',J}$, and the energy difference between the two levels, $\Delta \epsilon_{J',J} = \epsilon_{J'} - \epsilon{J}$, we can compute the energy gained by the medium per unit of volume and time through molecular thermal influence: 
\begin{equation}
    \Gamma_{\rm mol} = \sum_{J} \sum_{J'} n_{\rm mol}\, X_J\, B_{J,J'}\, \rho_{J,J'} P^c_{J',J}\, \Delta \epsilon_{J',J}.
    \label{eq:heating}
\end{equation}
We summed over all the $J'$ levels reachable by excitation as well as over $J$ to take into account the contribution from all the rotational levels.

The energy of a rotational level, $J$, can be computed as: 
 \begin{equation}
 \epsilon_J = h \, B J(J+1) \ {\rm where} \ B \ {\rm is \ the \ rotation \ constant} \ B=\frac{h}{8\pi^2 c I},
 \end{equation}
 with I the momentum of inertia. 

\subsection{Molecular cooling}
By analogy, we first compute the radiative de-excitation probability:  
\begin{equation}
    P^r_{J',J} =\frac{ X_{J'}A_{J',J} + X_{J'}B_{J',J}\rho_{J',J} } {n_\xi X_{J'}k^\xi_{J',J} + X_{J'}A_{J',J} + X_{J'}B_{J',J}\rho_{J',J}},
    \label{eq:radprob}
\end{equation}
which is the ratio of both induced and spontaneous radiative de-excitation rates, to the total one.

We change the radiative excitation probability from  Eq.~\ref{eq:heating} to the collisional excitation probability  with species $\xi$ : $n_\xi k^\xi_{J,J'}$ and change the collisional de-excitation probability to the radiative one (Eq.~\ref{eq:radprob}). We can then compute the energy lost by the medium per unit of volume and time through molecular thermal influence: 
\begin{equation}
    \Lambda_{\rm mol} = \sum_{J} \sum_{J'} n_{\rm mol}\, n_\xi\, k^\xi_{J,J'}\, X_J \,P^r_{J',J}\,\Delta \epsilon_{J',J}?
    \label{eq:cooling}
.\end{equation}
\vspace{0.5cm}

\newpage
\section{Molecular rotational levels}
\label{app:molpoplevels} 
This appendix is mainly inspired by ~\cite{puy97} and ~\cite{vonlthese}. The aim is to compute, $X_J$, the population fraction of molecules in each rotational level, $J$. \\

\subsection{Einstein coefficients}
\label{ein}

Let us assume a container inside which there is a radiation field of density $\mathrm{\rho(\nu)d\nu}$ and molecules capable of passing from a rotational level $J$ to a higher level $J'$ by the absorption of a photon whose energy $h_{\mathrm{P}} \nu_{JJ'}$ corresponds to the energy difference between the two levels (radiative excitation) or by collision (collisional excitation). De-excitation can also be collisional or radiative (spontaneous or induced).

There are three radiative processes: spontaneous emission, induced emission, and absorption. The Einstein coefficients $A_{J'J}$ and $B_{JJ'}$ are defined as follows:
\begin{itemize}
\item[$\bullet$] \texttt{Spontaneous emission:} the probability of spontaneous emission $A_{J'J}$ is the probability that a molecule in the rotational state $J'$ spontaneously emits a photon of energy $h_{\rm{P}} \nu_{JJ'}$ corresponding to the transition from $J'$ to $J$.
\item[$\bullet$] \texttt{Induced emission:} The probability of induced emission is defined by the probability that the atom passes from $J'$ to $J$ when it is subjected to induced radiation with a frequency between $\nu_{J'}$ and $\nu_{J'} + \mathrm{d}\nu$. This probability is denoted $B_{J'J} \rho(\nu)$.
\item[$\bullet$] \texttt{Absorption: } absorption occurs when a molecule in the $J$ state, exposed to isotropic radiation of density $\mathrm{\rho(\nu) d\nu}$ and frequency between $\nu_{J'}$ and $\nu_{J'} + \mathrm{d}\nu$, absorbs a photon of energy $h_{\rm{P}} \nu_{JJ'}$ and changes to the $J'$ state. We write the probability of absorption $B_{JJ'} \rho(\nu)$.
\end{itemize}

If we write $I$ the moment of inertia of a molecule, then the energy of the rotational level $J$ is given by \cite{cohen73} and \cite{kutner84}:

\begin{equation}
\epsilon_J = \frac{J(J+1) h_{\mathrm{P}}^2}{8 \pi^2 I},
\end{equation}
with $h_{\rm{P}}$ the Planck constant. We can rewrite this as:\  
\begin{equation}
\epsilon_J = h_{\mathrm{P}}BJ(J+1),
\end{equation}
where  $B = h_{\mathrm{P}}/8 \pi^2 I$ is called the molecule's rotation constant. We can see that the more massive a molecule is, the greater its moment of inertia will be, and therefore the smaller its rotation constant will be. Transition frequencies are obtained by taking the energy differences between two levels and dividing by $h_{\mathrm{P}}$. For dipole molecules, transitions obey the rule $\left| \Delta J \right| =1$, so that $J' = J+1$. In this case we have:
\begin{equation}
\nu_{J+1,J} = \nu_{J+1} - \nu_J = \frac{\epsilon_{J+1} - \epsilon_J}{h_{\mathrm{P}}} = 2B(J+1).
\end{equation}
For homopolar molecules like H$_2$, the transition rule is  $\left| \Delta J \right| =2$. Hence we get  
\begin{equation}
\nu_{J+2,J} = \nu_{J+2} - \nu_J = \frac{\epsilon_{J+2} - \epsilon_J}{h_{\mathrm{P}}} = 2B(2J+3).
\end{equation}
The Einstein coefficient, $A_{J+1,J}$, can be related to the dipole moment, $\mu_{\mathrm{r}}$~\citep{kutner84}:
\begin{equation}
A_{J+1,J} = \frac{64 \pi^4 \nu_{J+1,J}^3}{3h_{\mathrm{P}} c^3} \mu_{\mathrm{r}}^2 \frac{J+1}{2J+3}.
\label{coeffA}
\end{equation}
Furthermore, the coefficients  $B_{J+1,J}$ and  $A_{J+1,J}$  are related by the following formula:
\begin{equation}
B_{J+1,J} = A_{J+1,J} \frac{c^3}{8\pi h_{\mathrm{P}} \nu_{J+1,J}^3}.
\label{BfctA}
\end{equation}
Finally, the probabilities of absorption and stimulated emission are also related:
\begin{equation}
g_J B_{J,J+1} = g_{J+1} B_{J+1,J},
\label{BfctB}
\end{equation}
where the $g_i$ are the statistical weights of the rotational levels $i$: $g_i = 2i + 1$.\\

In the case of collisional transitions, the de-excitation probability $C_{J',J}$ is computed by the product:
\begin{equation}
C_{J',J} = \tau_{\mathrm{coll}} n_{\mathrm{coll}},
\end{equation}
where $\tau_{\mathrm{coll}}$ is the collision rate between the molecule and the collisional specie whose density is denoted by  $n_{\mathrm{coll}}$.  Assuming a Maxwellian velocity distribution, the collisional excitation probability $C_{J,J'}$ is related to $C_{J',J}$ by
\begin{equation}
C_{J,J'} = C_{J',J} \frac{g_{J'}}{g_J} \exp \left( h_{\mathrm{P}} \nu_{J+1,J} / k_{\mathrm{B}} T_{\mathrm{m}} \right).
\label{coeffC}
\end{equation}

From the above considerations we conclude that two quantities mainly determine the efficiency of the $\Psi$ thermal function of a molecular species: the dipole moment $\mu_{\mathrm{r}}$ and the rotation constant $B$. Since the Einstein coefficients depend on the square of the dipole moment, molecules for which $\mu_{\mathrm{r}}$ is large will be very efficient thermal agents. However, of the two most abundant molecules in the cosmic gas, H$_2$ does not have a permanent dipole moment, and the one of HD is very weak. For H$_2$ , rotational constant is B$_{{\rm H}_2} =85.35$ K with no dipole moment $\mu_r ({\rm H_2}) =0$ D, while for HD  we have B$_{\rm HD}=64.30$ K and the dipole moment $\mu_r ({\rm HD})= 8.3 \times 10^{-4}$ D.

\subsection{Populations of the rotational levels}

To calculate the thermal function of a molecule, we need to know the populations of its rotational levels $X_J$. When molecules have permanent electric dipole moments (along the axis of the molecule), the strongest transitions are those that obey the selection rule $|\Delta J| = 1$. In the case of homopolar molecules such as molecular hydrogen, the dipole moment is zero for symmetry reasons. Transitions between rotational levels are therefore quadrupolar. This type of transition imposes the rule $|\Delta J| = 2$. In this case, transitions between even-numbered levels $J$ = 0, 2, 4, \ldots \ are called para transitions. Those occurring between odd levels $J$ = 1, 3, 5, \ldots \ are called ortho transitions. We consider 20 rotational levels in our calculations.

\paragraph{\textbf{H$_2$ Molecule}}

\label{molhomo}
$\mathrm{H_2}$ is the only primordial homopolar molecule. Its rotational constant is equal to $B_{\mathrm{H_2}} = 170.66$ K. The radiative transition probability is given by 
\begin{equation}
A_{J,J-2} = 7.52 \cdot 10^{-13} \frac{J(J-1)(2J-1)^4}{2J+1} \quad \mathrm{s}^{-1}.
\label{AEinsteinSpitzer}
\end{equation}
Let's consider three consecutive rotational levels  $J-2$, $J$ and $J+2$. The variation of the $J$ level population is :
\begin{equation}
\begin{array}{ccl}
\frac{\mathrm{d}X_J}{\mathrm{d}t} & = & \underbrace{C_{J-2,J}X_{J-2} + B_{J-2,J}\rho_{J-2,J}X_{J-2} + C_{J+2,J}X_{J+2} + B_{J+2,J}\rho_{J+2,J}X_{J+2} + A_{J+2,J}X_{J+2}}_{\mathrm{Gain}}\\
& - & \underbrace{\left( C_{J,J+2}X_J + B_{J,J+2}\rho_{J,J+2}X_J + C_{J,J-2}X_J + B_{J,J-2}\rho_{J,J-2}X_J + A_{J,J-2}X_J \right)}_{\mathrm{Loss}},
\label{dxjdt}
\end{array}
\end{equation}
The first two terms of the second member in the RHS  represent the transition from level $J$ to level $J+2$. The last three terms express the transition from $J$ to $J-2$. The $J$ level can therefore be depopulated either by transition to the $J+2$ level or by de-excitation to the lower $J-2$ level. The transitions are almost instantaneous, leading to a pseudo-stationary state $\frac{{\rm d}X_J}{\mathrm{dt}} = 0$. After rearranging the terms we get:

\begin{equation}
\begin{array}{cl}
& C_{J+2,J}X_{J+2} + A_{J+2,J}X_{J+2} + B_{J+2,J}\rho_{J+2,J}X_{J+2} - C_{J,J+2}X_J - B_{J,J+2}\rho_{J,J+2}X_J\\
= & C_{J,J-2}X_J + A_{J,J-2}X_J + B_{J,J-2}\rho_{J,J-2}X_J - C_{J-2,J}X_{J-2} - B_{J-2,J}\rho_{J-2,J}X_{J-2}.
\label{xj}
\end{array}
\end{equation}
The symmetry of this last expression for the coefficients $J-2$, $J$ and $J+2$ allows us to conclude that the coefficients $J$ are independent for each member of the equality. We then obtain:
\begin{equation}
C_{J+2,J}X_{J+2} + A_{J+2,J}X_{J+2} + B_{J+2,J}\rho_{J+2,J}X_{J+2} - C_{J,J+2}X_J - B_{J,J+2}\rho_{J,J+2}X_J = \mathrm{cst}.
\label{xjxj+2}
\end{equation}
Through recurrence, we can get back to the fundamental level $J = 0$:
\begin{equation}
C_{2,0}X_2 + A_{2,0}X_2 + B_{2,0}\rho_{2,0}X_{2} - C_{0,2}X_0 - B_{0,2}\rho_{0,2}X_0 = \mathrm{cst}.
\label{cte}
\end{equation}
And according to Eq.~\ref{dxjdt}, we also have the pseudo-stationarity hypothesis:
\begin{equation}
0 = \frac{d\mathrm{X_0}}{\mathrm{d}t} = C_{2,0}X_2 + A_{2,0}X_2 + B_{2,0}\rho_{2,0}X_{2} - C_{0,2}X_0 - B_{0,2}\rho_{0,2}X_0.
\end{equation}
Comparing these two expressions, we conclude that the constant in Eq.~\ref{cte} is null. Equation~\ref{xjxj+2} leads to a simple relation between the populations $X_J$ and $X_{J+2}$:
\begin{equation}
X_J = X_{J+2} \frac{C_{J+2,J} + A_{J+2,J} + B_{J+2,J}\rho_{J+2,J}}{C_{J,J+2} + B_{J,J+2}\rho_{J,J+2}}.
\label{relxj}
\end{equation}
In addition, the energy density of cosmological blackbody radiation at the transition frequency $\nu_{J+2,J}$ is given by
\begin{equation}
\rho_{J+2,J} = \rho_{J,J+2} = \frac{8\pi h_{\mathrm{P}} \nu_{J+2,J}^3}{c^3} \frac{1}{\exp \left( \frac{h_{\mathrm{P}} \nu_{J+2,J}}{k_{\mathrm{B}} T_{\mathrm{\gamma}}} \right) -1}.
\end{equation}
This relation helps us to simplify Eq.~\ref{relxj}:
\begin{equation}
X_{J+2} = X_J \frac{C_{J,J+2} \; \left( \exp \left( \frac{T_{J,J+2}}{T_{\mathrm{\gamma}}} \right) - 1 \right) + A_{J+2,J} \; \frac{2J+5}{2J+1}}{C_{J+2,J} \; \left( \exp \left( \frac{T_{J,J+2}}{T_{\mathrm{\gamma}}} \right) - 1 \right) + A_{J+2,J} \; \exp \left( \frac{T_{J,J+2}}{T_{\mathrm{\gamma}}} \right)},
\end{equation}
Where we introduced the transition temperature $T_{J,J+2}$ such that $k_{\mathrm{B}} T_{J,J+2} = h_{\mathrm{P}} \nu_{J,J+2}$. With the relation (Eq.~\ref{coeffC}) between the collision probabilities, we get to the following expression:
\begin{equation}
X_{J+2} = X_J \frac{C_{J,J+2} \; \left( \exp \left( \frac{T_{J,J+2}}{T_{\mathrm{\gamma}}} \right) - 1 \right) + A_{J+2,J} \; \frac{2J+5}{2J+1}}{C_{J,J+2} \; \frac{2J+5}{2J+1} \; \exp \left( \frac{T_{J+2,J}}{T_{\mathrm{m}}} \right) \; \left( \exp \left( \frac{T_{J,J+2}} {T_{\mathrm{\gamma}}} \right) - 1 \right) + A_{J+2,J} \; \exp \left( \frac{T_{J,J+2}}{T_{\mathrm{\gamma}}} \right)}.
\end{equation}
We introduce the probability  $X_{J+2} = a_{J+2} X_J$, where :
\begin{equation}
a_{J+2} := \frac{C_{J,J+2} \; \left( \exp \left( \frac{T_{J,J+2}}{T_{\mathrm{\gamma}}} \right) - 1 \right) + A_{J+2,J} \; \frac{2J+5}{2J+1}}{C_{J,J+2} \; \frac{2J+5}{2J+1} \; \exp \left( \frac{T_{J+2,J}}{T_{\mathrm{m}}} \right) \; \left( \exp \left( \frac{T_{J,J+2}}{T_{\mathrm{\gamma}}} \right) - 1 \right) + A_{J+2,J} \; \exp \left( \frac{T_{J,J+2}}{T_{\mathrm{\gamma}}} \right)}.
\end{equation}
This gives us the formulas for calculating the $J$ level population :
\begin{equation}
\begin{array}{ccll}
X_{2n} & = & \left( \prod\limits_{i=1}^{n} a_{2i} \right) X_0 & \mathrm{for \; para \; transitions},\\
X_{2n+1} & = & \left( \prod\limits_{i=1}^{n} a_{2i+1} \right) X_1 & \mathrm{for \; ortho \; transitions}.
\end{array}
\end{equation}
Considering that the sum over the populations of all the rotational levels is normalized, the final result for the rotational levels populations is the following : 
\begin{eqnarray}
X_{\mathrm{0}} & = & \frac{1}{1 + \sum_{n=1}^{\infty} \left( \prod_{i=1}^{n} a_{2i} \right)},\\
X_{\mathrm{1}} & = & \frac{1}{1 + \sum_{n=1}^{\infty} \left( \prod_{i=1}^{n} a_{2i+1} \right)},\\
X_{2n} & = & \frac{\prod_{i=1}^{n} a_{2i}}{1 + \sum_{n=1}^{\infty} \left( \prod_{i=1}^{n} a_{2i} \right)},\\
X_{2n+1} & = & \frac{\prod_{i=1}^{n} a_{2i +1}}{1 + \sum_{n=1}^{\infty} \left( \prod_{i=1}^{n} a_{2i + 1} \right)},
\end{eqnarray}
where $n$ is a non-zero integer.

\paragraph{\textbf{HD molecule}}

\label{moldip}
A similar computation to the homopolar case gives the following results for the rotational level populations:

\begin{eqnarray}
X_{\mathrm{0}} & = & \frac{1}{1 + \sum_{J=1}^{\infty} \left( \prod_{i=1}^{J} a_i \right)},\\
X_J & = & \frac{\prod_{i=1}^{J} a_i}{1 + \sum_{J=1}^{\infty} \left( \prod_{i=1}^{J} a_i \right)}, J \neq 0.
\end{eqnarray}
The proportionality factors are now: 
\begin{eqnarray}
a_{J+1} & = & \left[ C_{J,J+1} \; \left( \exp \left( \frac{T_{J,J+1}}{T_{\mathrm{\gamma}}} \right) - 1 \right) + A_{J+1,J} \; \frac{2J+3}{2J+1}\right] \nonumber \\
& \times & \left[ C_{J,J+1} \frac{2J+3}{2J+1} \; \exp \left( \frac{T_{J+1,J}}{T_{\mathrm{m}}} \right) \; \left( \exp \left( \frac{T_{J,J+1}} {T_{\mathrm{\gamma}}} \right) - 1 \right) \right. \nonumber \\
& + & \left. A_{J+1,J} \; \exp \left( \frac{T_{J,J+1}}{T_{\mathrm{\gamma}}} \right) \right]^{-1}.
\end{eqnarray}

\end{appendix}
\setcounter{figure}{0}
\setcounter{equation}{0}
\setcounter{section}{2}

\end{document}